\documentclass[12pt,preprint]{aastex}

\slugcomment{Not to appear in Nonlearned J., 45.}

\shorttitle{Acretion Rates of PMS stars in the LMC}
\shortauthors{Romaniello et al.}

\newcommand{\ha}{\ensuremath{\mathrm{H}\alpha}}
\newcommand{\ie}{\emph{i.e.} }
\newcommand{\ebv}{\ensuremath{\mathrm{E}(\mathrm{B}-\mathrm{V})}}

\begin{document}

\title{Low-Mass Pre-Main Sequence Stars in the Large Magellanic Cloud - III:
Accretion Rates from HST-WFPC2 Observations$^\dag$}
\renewcommand{\thefootnote}{\fnsymbol{footnote}}\footnotetext[2]{Based
on observations with the NASA/ESA Hubble Space Telescope, obtained at
the Space Telescope Science Institute, which is operated by
AURA,~Inc., under NASA contract
NAS~5-26555.}\renewcommand{\thefootnote}{\arabic{footnote}}

\author{M. Romaniello}
\affil{European Southern Observatory, Karl-Schwarzschild-Strasse 2,
D-85748 Garching bei M\"{u}nchen, Germany}
\email{mromanie@eso.org}

\and

\author{M. Robberto\altaffilmark{1} and N. Panagia\altaffilmark{1}}
\affil{Space Telescope Science Institute, 3700 San Martin Drive, Baltimore,
MD 21218}
\email{robberto@stsci.edu, panagia@stsci.edu}

\altaffiltext{1}{Affiliated to the Astrophysics Division, Space Science
Department of ESA.}

\begin{abstract}
We have measured the present accretion rate of roughly 800 low-mass
($\sim1-1.4~M_\sun$) pre-Main Sequence stars in the field of
Supernova~1987A in the Large Magellanic Cloud (LMC, $Z\simeq 0.3\
Z_\sun$). It is the first time that this fundamental parameter for
star formation is determined for low-mass stars outside our Galaxy.
The Balmer continuum emission used to derive the accretion rate
positively correlates with the \ha\ excess. Both these phenomena are
believed to originate from accretion from a circumstellar disk so that
their simultaneous detection provides an important confirmation of the
pre-Main Sequence nature of the \ha\ and UV excess objects, which are
likely to be the LMC equivalent of Galactic Classical T~Tauri
stars. The stars with statistically significant excesses are measured
to have accretion rates larger than $\sim 1.5\times10^{-8}M_\sun\
yr^{-1}$ at an age of 12-16~Myrs. For comparison, the time scale for
disk dissipation observed in the Galaxy is of the order of
6~Myrs. Moreover, the oldest Classical T~Tauri star known in the Milky Way
(TW~Hydr\ae, with 10~Myrs of age) has a measured accretion rate of
only $5\times 10^{-10}~M_\sun/yr$, \ie 30 times less than what we
measure for stars at a comparable age in the LMC. Our findings
indicate that metallicity plays a major role in regulating the
formation of low-mass stars.
\end{abstract}


\section{Introduction\label{sec:intro}}
The processes at play during star formation determine much of the
appearance of the visible Universe. The shape of the stellar Initial
Mass Function (IMF) and its normalization (the star-formation rate)
are, together with stellar evolution theory, key ingredients in
determining the chemical evolution of a galaxy and its stellar
content. Yet, our theoretical understanding of the processes that lead
from diffuse molecular clouds to stars is still very tentative, as
many complex physical phenomena concur in producing the final
results. While clear variations in the star-formation rate are
observed in different regions of the Milky Way and in external
galaxies, with their histories showing bursts and lulls
\citep[e.g.,][]{tol00}, the observational evidence for (or against)
variations in the IMF is often contradictory \citep[see the review of]
[or the Gilmore vs Eisenhauer debate in ``Starbursts: Near and Far',
2001]{scalo98}. Yet, variations in the IMF can dramatically alter the
chemical evolution of a galaxy \citep[e.g.,][]{wyse98}.

From an observational standpoint, most of the effort has traditionally
been devoted to nearby Galactic star-forming regions, such as the
\objectname{Taurus-Auriga} complex, \objectname{Orion}, etc. If this,
on the one hand, permits one to observe very faint stars at the best
possible angular resolution, on the other it is achieved at the
expense of probing only a very limited set of initial conditions for
star formation \citep[all these clouds have essentially solar
metallicity, e.g.,][]{padg96}.

Studying the effects of a lower metallicity on star formation is also
essential to understand the evolution of both our own Galaxy, in which
a large fraction of stars were formed at metallicities below solar,
and what is observed at high redshifts. As a matter of fact, the
global star formation rate appears to have been much more vigorous (a
factor of 10 or so) at $z\simeq 1.5$ than it is today \citep[][and
subsequent incarnations of the so-called ``Madau plot'']{mad96}.  At
that epoch the mean metallicity of the interstellar gas was similar to
that of the \objectname{Large Magellanic Cloud} at present \citep[LMC,
e.g.][]{pei99}. This fact makes the study of star forming regions in
the LMC especially important for the understanding of galaxy
evolution.

With a distance modulus of $18.57\pm0.05$ \citep[see the discussion
in][]{rom00}, the LMC is our closest galactic companion after the
\objectname{Sagittarius} dwarf galaxy. At this distance one arcminute
corresponds to about 15~pc and, thus, one pointing with a typical
imaging instrument comfortably covers almost any star forming region
in the LMC \citep[10~pc see, e.g.,][]{hod88}. In particular, the field
of view of about $2.7\arcmin\times2.7\arcmin$ of the WFPC2 on board
the HST corresponds to $37~\mathrm{pc}\times37~\mathrm{pc}$, and leads
to the detection of several thousands of stars per pointing
\citep[e.g., ][]{rom02}. The LMC is especially suited for stellar
populations studies for two additional reasons. First, the depth of
the LMC along the line of sight is negligible, at least in the central
parts we consider \citep{mar01}. All of the stars can, then,
effectively be considered at the same distance, thus eliminating a
possible spurious scatter in the Color-Magnitude Diagrams. Second, the
extinction in its direction due to dust in our Galaxy is low, about
$\ebv\simeq 0.05$ \citep{bes91,sch91} and, hence, our view is not
severely obstructed.

There is currently a widespread agreement that low mass stars form by
accretion of material until their final masses are reached
\citep[e.g.][and references therein]{bon01}. As a consequence, the
accretion rate is arguably \emph{the} single most important parameter
governing the process of low-mass star formation and its final
results, including the stellar Initial Mass Function.  Ground and
HST-based studies show that there may be significant differences
between star formation processes in the LMC and in the Galaxy. For
example, \citet{lam99} and \citet{dewit02} have identified by means of
ground-based observations high-mass pre-Main Sequence stars (Herbig
AeBe stars) with luminosities systematically higher than observed in
our Galaxy, and located well above the ``birthline'' of
\citet{ps91}. They attribute this finding either to a shorter
accretion timescale in the LMC or to its smaller dust-to-gas
ratio. Whether such differences in the physical conditions under which
stars form will generally lead to differences at the low mass end is
an open question, but \citet{pan00} offer tantalizing evidence of a
higher accretion also for LMC low mass stars.

In this paper we present the first measurement of the accretion rate
onto low-mass pre-Main Sequence stars outside of our Galaxy. The data
and the reduction process are presented in the next section, while the
detection of Balmer continuum excess and its conversion into an
accretion rate are described in
section~\ref{sec:measuring_acc}. Section~\ref{sec:pms_age} is devoted
to deriving the stellar parameters of our sample, \ie the stars'
masses and ages. Finally, the conclusions are drawn in
section~\ref{sec:sum}.

\section{Observation and data reduction\label{sec:obs}}
The field of \objectname{SN~1987A} in the LMC was repeatedly imaged
over the years with the WFPC2 on-board the HST to monitor the
evolution of its Supernova remnant. We have taken advantage of this
wealth of data and selected from the HST archive a uniform dataset
providing broad-band coverage from the ultraviolet to the near
infrared, as well as imaging in the \ha\ line. The log of the
observations we have used is reported in Table~\ref{tab:log}. A
description of the camera and its filter set can be found in
\citet{bir01}. All images are centered with the Planetary Camera chip
on SN~1987A ($\alpha_{2000}=05:35:28.26, \delta_{2000}=-69:16:13.0$),
but have different position angles on the sky, resulting in complete
coverage of an almost circular region of $130\arcsec$ (about 30~pc) in
radius.

The data were processed through the standard Post Observation Data
Processing System pipeline for bias removal and flat fielding.  In all
cases cosmic ray events were removed combining the available images
after accurate registration and alignment.

The plate scale is 0.045 and 0.099 arcsec/pixel in the Planetary
Camera and in the three Wide Field chips, respectively. We performed
aperture photometry following the prescriptions by \citet{gil90} as
refined by \citet{rom98}, \ie measuring the flux in a circular
aperture of 2~pixels radius and the sky background value in an annulus
of internal radius 3~pixels and width 2~pixels. Due to the
undersampling of the WFPC2 Point Spread Function, this prescription
leads to a smaller dispersion in the Color-Magnitude Diagrams, \ie
better photometry, than PSF fitting for non-jittered observations of
marginally crowded fields \citep{coo95,rom98}.  Photometry for the
saturated stars was recovered by either fitting the unsaturated wings
of the PSF for stars with no saturation outside the central 2 pixel
radius, or by following the method developed by \citet{gil94} for the
heavily saturated ones. The flux calibration was done using the
internal calibration of the WFPC2 \citep{whit95}, which is typically
accurate to within 5\% at optical wavelengths. The spectrum of Vega is
used to set the photometric zeropoints (VEGAMAG system).

In the following, as a measure of the overall photometric accuracy, we will
use the mean error in five broad bands defined as:

\begin{equation}
  \bar{\delta}_5=\sqrt{\frac{\delta(m_\mathrm{F336W})^2+
  \delta(m_\mathrm{F439W})^2+\delta(m_\mathrm{F555W})^2+
  \delta(m_\mathrm{F675W})^2+\delta(m_\mathrm{F814W})^2}{5}}
  \label{eq:e2m}
\end{equation}

\subsection{From colors to luminosity and temperature\label{sec:dered}}
Once the observed fluxes of the stars are carefully measured, we
derive their intrinsic properties, \ie luminosity and temperature, as
well as the extinction caused by the intervening interstellar dust
along the line of sight, using the prescriptions developed by
\citet{rom02}. The intrinsic stellar parameters and their associated
errors are derived with a minimum $\chi^2$ technique by comparing the
observed magnitudes to the ones expected based on the theoretical
stellar atmosphere models of \citet{bes98} computed in the HST-WFPC2
bands using the IRAF \emph{synphot} task. In order to cope with the
effects of interstellar dust we have used the the extinction law
appropriate for this region of the LMC \citep{scud96}. Let us stress
here that, by convolving the theoretical spectra with the filter
sensitivity curves provided in IRAF, we ensure that the observations
are faithfully modeled. In particular, the red leak that affects the
WFPC2 F336W (U-band-like) filter is properly taken into account.

The dereddening method is extensively described in \citet{rom02}, but
let us briefly summarize it here:

\begin{enumerate}
  \item stars for which both $E(B-V)$ and $T_{eff}$ can be
     simultaneously derived are selected according to their
     photometric error ($\bar{\delta}_5<0.1$) and their location in 
     the $Q_{UBI}$ vs $(U-I)$ plane, where $Q_{UBI}$ is a reddening-free
     color defined as:

     \begin{displaymath}
       Q_{UBI}\equiv (U-B)-\frac{E(U-B)}{E(B-I)}\ (B-I)
     \end{displaymath}

     This color-based selection is aimed at solving the possible
     non-monotonicity of broad-band colors with temperature
     \citep[see, for example,][p207]{allen73}. The stars for which
     $E(B-V)$ and $T_{eff}$ can be derived simultaneously turn out to
     be hotter than 10,000~K or between 6,750 and 8,500~K.

     Also, a star's location in the $Q_{UBI}$ vs $(U-I)$ plane
     provides a first guess of its temperature and reddening;

  \item for the stars selected in step 1, $E(B-V)$, $T_{eff}$, $L$ and
     their associated uncertainties are derived starting from the
     first guesses by performing a minimum $\chi^2$ multi-band fit of
     synthetic colors from \citet{bes98} to the observed magnitudes;

  \item for each star for which $E(B-V)$ and $T_{eff}$ cannot be
     derived simultaneously because of its intrinsic temperature
     and/or too large errors, the reddening is set as the mean of the
     ones of its 4 closest neighbors with direct reddening
     determination. The corresponding rms is used as an estimate of
     the uncertainty on the adopted value of $E(B-V)$. The effective
     temperature, luminosity and associated errors are, then, derived
     from a minimum $\chi^2$ multi-band fit.

     In the case of the field discussed here, there is, on average,
     one star with direct $E(B-V)$ determination every 13
     arcsec$^2$. Of course, it is possible that a few stars have, in
     reality, extinction values significantly different from the local
     mean, but the global effect is negligible.
\end{enumerate}

The errors on $E(B-V)$ and $T_{eff}$ are computed from the $\chi^2$
maps and propagated to the luminosity. A full discussion on the errors
is reported in \citet{rom02}, but it is important to keep in mind here
that the procedure outlined above does not introduce any systematic
errors on the derived stellar parameters.

\citet{pan00}, using this same dataset, have measured the age of the
massive stars in the field to be $12\pm2$~Myrs. Also, they have
identified several hundreds pre-Main Sequence stars through their \ha\
emission. We will now measure, for the first time outside our Galaxy,
the accretion rate on these low mass pre-Main Sequence stars.

\section{Measuring the accretion rate\label{sec:measuring_acc}}
The idea that the strong excess emission observed in some Galactic
low-mass, pre-Main Sequence stars (T~Tauri stars) is produced by
accretion of material from a circumstellar disk dates back to the
pioneering work of \citet{lynd74}. The excess luminosity is, then,
related to the mass accretion rate. In particular, the Balmer
continuum radiation produced by the material from the disk as it hits
the stellar surface has been used as an estimator of the mass infall
activity \citep[see, for example,][and references therein]{gull98}.

\subsection{The U-band excess\label{sec:uex}}
\begin{figure}
  \plotone{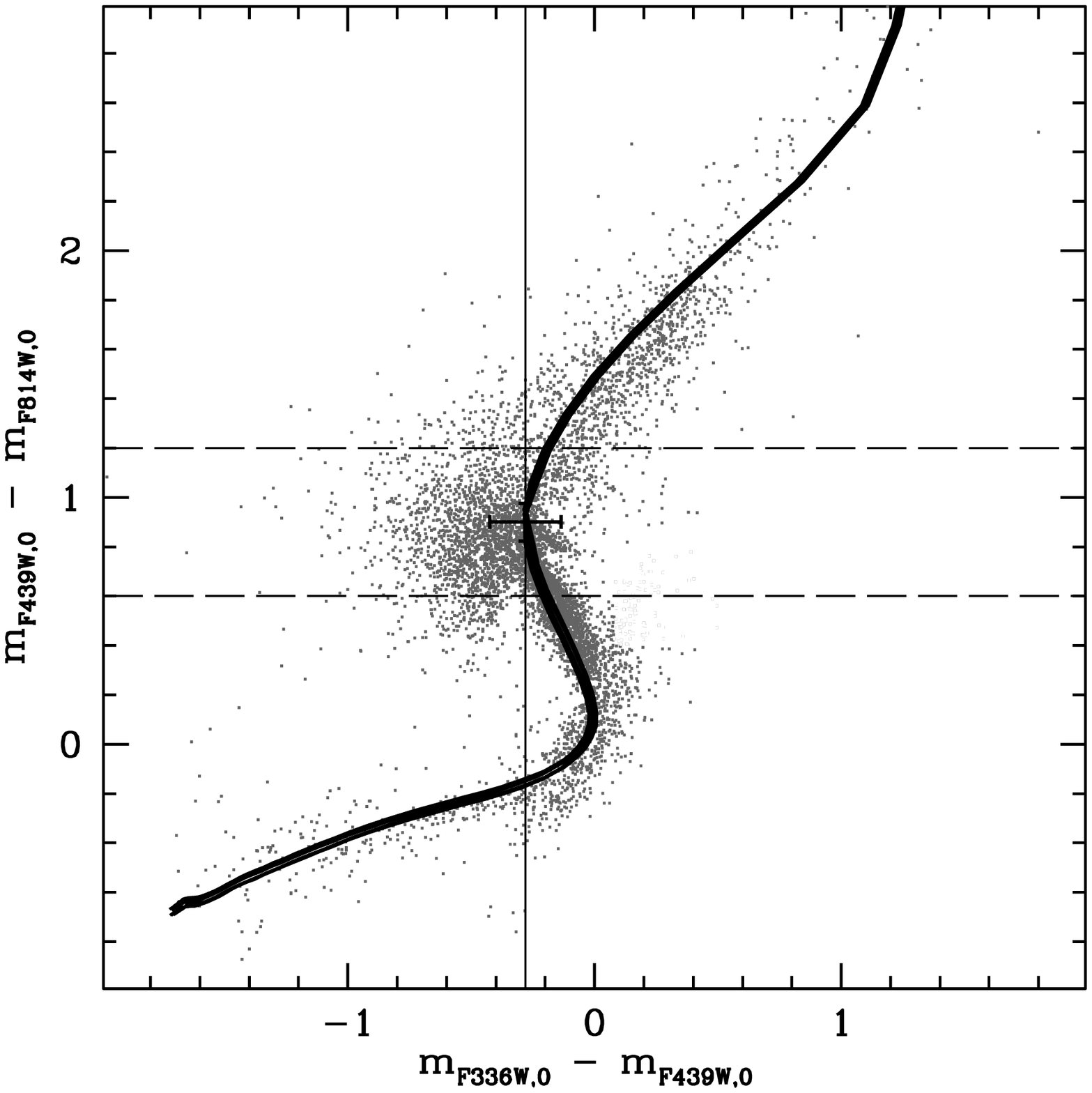}
  \caption{Locus in the $(m_\mathrm{F336W,0}-m_\mathrm{F439W,0})$ vs
    $(m_\mathrm{F439W,0}-m_\mathrm{F814W,0})$ plane of the stars with
    good overall photometry, \ie a mean error in the broad bands
    between F336W and F439W smaller than 0.1~mag (see
    equation~\ref{eq:e2m}). The theoretical locus from the models of
    \citet{bes98} for $Z=0.3\ Z_\sun$ and $\log(g)=4,5$ is shown
    as a full line. The thin vertical line marks the position of the
    peak of the histogram in Figure~\ref{fig:hubex}. The horizontal
    dashed lines highlight the region used to select stars with excess
    F336W emission ($0.6<(m_\mathrm{F439W,0}-m_\mathrm{F814W,0})<1.2$,
    see text). The presence of stars with excess F336W emission is
    evident inside this box at
    $(m_\mathrm{F439W,0}-m_\mathrm{F814W,0})\simeq0.9$. The typical
    uncertainties for these stars is shown by the
    errorbar. \label{fig:ubbi}}
\end{figure}

The presence of a rich population of stars with U-band excess in the
field of SN~1987A is clearly detected in Figure~\ref{fig:ubbi}, where
we plot as gray dots the dereddened
$(m_\mathrm{F336W,0}-m_\mathrm{F439W,0})$ vs
$(m_\mathrm{F439W,0}-m_\mathrm{F814W,0})$, \ie
$(\mathrm{U}_0-\mathrm{B}_0)$ vs $(\mathrm{B}_0-\mathrm{I}_0)$, colors
for stars with good overall photometry ($\bar{\delta}_5<0.1$). The
F336W and F439W filters bracket the Balmer jump and an excess in the
$(m_\mathrm{F336W,0}-m_\mathrm{F439W,0})$ color translates in an
excess emission in the Balmer continuum. The typical errorbar,
computed as the mean of the photometric errors of the stars with
$0.6<(m_\mathrm{F439W,0}-m_\mathrm{F814W,0})<1.2$, is shown as a
cross. For reference, the expected locus from the stellar atmosphere
models of \citet{bes98} for $Z=0.3\ Z_\sun$ and $\log(g)=4.5$ is
shown as a solid line. At the end of this section we will discuss the
influence of chemical composition and surface gravity on our results.

The reddening and intrinsic stellar parameters for the stars with
possible U band excess were computed excluding the F336W magnitudes
from the multi-band minimum $\chi^2$ fit described in
section~\ref{sec:dered}. This is because the excess itself can make
the observed F336W flux significantly different from that of a
photosphere, hence yielding incorrect results for the stellar
parameters. For example, the roughly 600 stars under the Main Sequence
noted by \citet{pan00} in this same field have too high a temperature
precisely because the U band was not excluded from the fit. As
detailed in \citet{rom02}, however, it is not possible to reliably
derive both $T_{eff}$ and $E(B-V)$ without the U band.  For the stars
with possible U band excess, then, we have used the reddening value
from their 4 nearest neighbors with a direct determination of
$E(B-V)$ and derived the stellar parameters accordingly.

The presence in Figure~\ref{fig:ubbi} of stars with too blue a
$(m_\mathrm{F336W,0}-m_\mathrm{F439W,0})$ color given the observed
$(m_\mathrm{F439W,0}-m_\mathrm{F814W,0})$ color is apparent at
$(m_\mathrm{F439W,0}-m_\mathrm{F814W,0})\simeq0.9$.  To quantify the
U-band excess we consider the distribution of the
$(m_\mathrm{F336W,0}-m_\mathrm{F439W,0})$ color as shown in
Figure~\ref{fig:hubex}. There, the solid line shows the color
distribution of the stars within the color interval
$0.6<(m_\mathrm{F439W,0}-m_\mathrm{F814W,0})<1.2$.  The FWHM expected
from the measured photometric and dereddening errors is shown by the
horizontal errorbar. It nicely reproduces the width of the right-hand
side of the observed histogram, while it severely underestimates the
one on the left (the dashed line is the histogram produced by the
reflection of the red part of the distribution relative to its peak
value). Hence, we take the resulting symmetric distribution, which has
a half-power width of about 0.25~magnitudes, as an indication of the
broadening caused by photometric and dereddening errors. This is
justified by noticing that the peak of the observed
$(m_\mathrm{F336W,0}-m_\mathrm{F439W,0})$ distribution occurs where
the models atmospheres predict it to occur
($(m_\mathrm{F336W,0}-m_\mathrm{F439W,0})=-0.27$, thin vertical line
in Figure~\ref{fig:ubbi}), which provides an independent check on the
dereddening procedure we have used, ruling out any significant
systematic errors on the inferred dereddened colors.  An excess of
stars with F336W emission as compared to a normal photosphere, \ie
blue $(m_\mathrm{F336W,0}-m_\mathrm{F439W,0})$ colors, is evident
(hatched area).

\begin{figure}
  \plotone{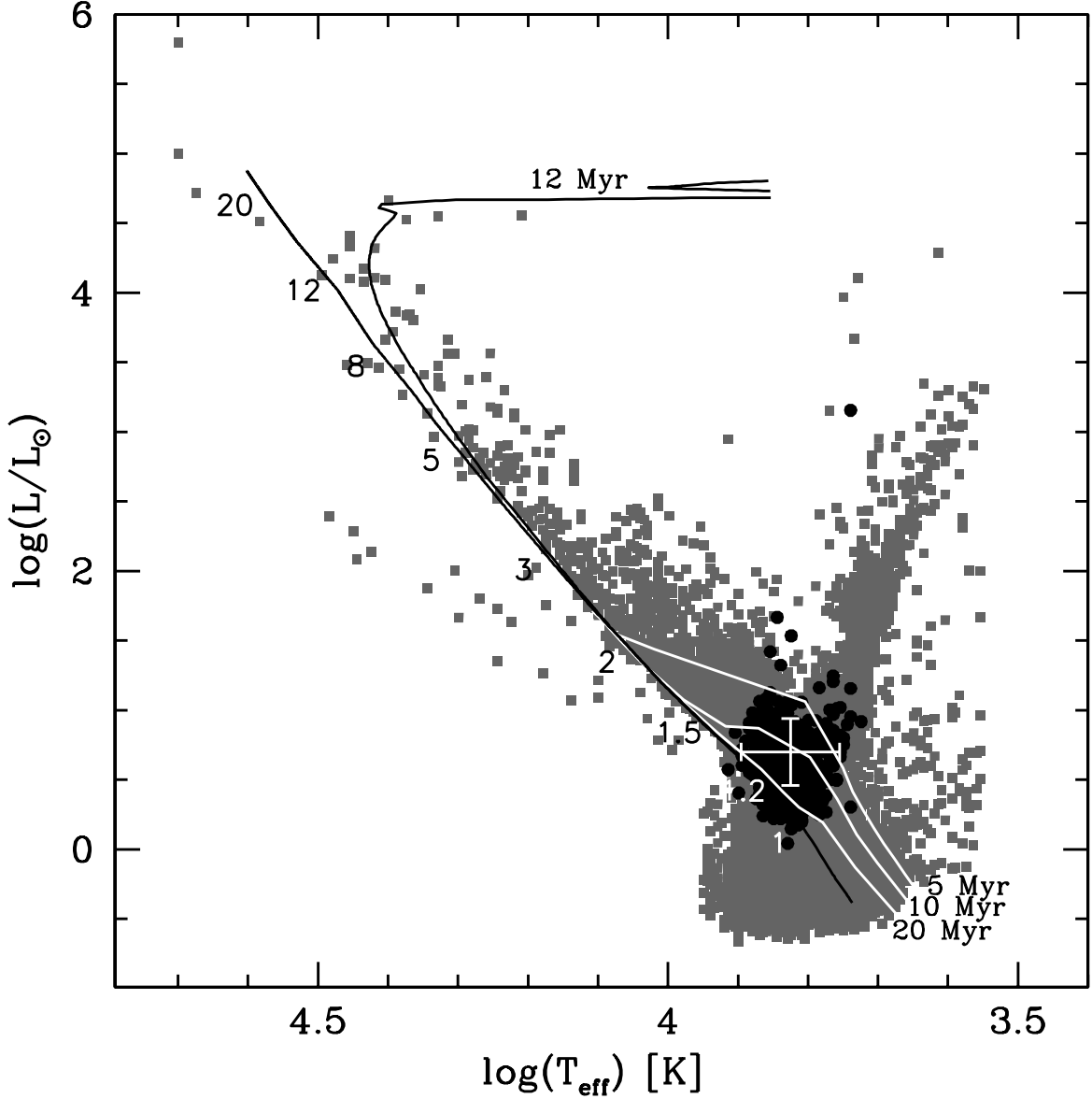}
  \caption{HR diagram displaying the position of the stars with U
   excess (black dots) overlaid on the general stellar population
   (gray squares) found in the WFPC2 field. Luminosities and
   temperatures for the stars with excess were computed excluding the
   F336W magnitude from the fit to the model atmospheres and adopting
   for each star the mean $E(B-V)$ value of its 4 closet neighbors.
   The typical uncertainties on the luminosity and temperature for the
   stars with U excess, computed as the mean of the uncertainties on
   the individual stars, is shown as a cross.  For reference, we also
   plot the theoretical Zero Age Main Sequence, with the position of
   stars of various masses marked on it, and a 12~Myr post-Main
   Sequence isochrone \citep[$Z=0.3\ Z_\sun$,][]{bro93}. Also
   shown are 5-20~Myrs pre-Main Sequence isochrones \citep{sie97},
   again computed for $Z=0.3\ Z_\sun$.\label{fig:hr_uvex}}
\end{figure}

The location in the HR diagram of the stars with and without U excess
is shown in Figure~\ref{fig:hr_uvex}. Also indicated are the
theoretical Zero Age Main Sequence and 12~Myr post-Main Sequence
isochrone from \citet{bro93} and the pre-Main Sequence isochrones by
\citet{sie97} for $Z=0.3\ Z_\sun$, the metallicity appropriate for the
young population of the Large Magellanic Cloud
\citep{hil95,geha98}. The typical uncertainties on the luminosity and
temperature for the stars with U excess, computed as the mean of the
uncertainties on the individual stars, is shown as a cross. If
interpreted as pre-Main Sequence stars, then, the position of the
stars with U excess in the HR diagram of Figure~\ref{fig:hr_uvex} is
consistent at a with an single age of 12~Myr, the age inferred from
the upper Main Sequence \citep[see also][]{pan00}. We will return on
the age dating of these stars in section~\ref{sec:pms_age}.

We can now take a step forward and compute the excess
\emph{luminosity} in the F336W filter, which, as we will see in
section~\ref{sec:uex2acc}, is directly related to the mass accretion
rate onto the pre-Main Sequence star. This is accomplished by
computing the difference between the measured dereddened flux in the
band and the one predicted by the stellar atmosphere models
\citep{bes98} for a photosphere with the effective temperature of the
star. We have derived this latter quantity by fitting the observed
BVRI magnitudes to the same model atmospheres of
\citet{bes98}:

\begin{equation}
  L_\mathrm{F336W,exc}=4\pi\ \Delta\lambda_\mathrm{F336W}\
  \left(D^2\ f_\mathrm{F336W,obs}-R_\ast^2\ F_\mathrm{F336W,mod}\right)
  \label{eq:ulex}
\end{equation}
where $\Delta\lambda_\mathrm{F336W}=370$~\AA\ is the width of the
F336W filter \citep{bir01}, $D=51.8$~kpc is the distance to this LMC
field \citep{rom00}, $R_\ast$ the stellar radius and
$f_\mathrm{F336W,obs}$ and $F_\mathrm{F336W,mod}$ are the observed and
model flux densities, respectively. The stellar radius, like the
effective temperature, is obtained by fitting the observed BVRI
magnitudes to the same model atmospheres of \citet{bes98}. The U
magnitude was excluded from the fit because it can be significantly
contaminated by the excess, hence yielding incorrect results for the
stellar parameters.

The distribution of the resulting excesses in solar units is shown in
Figure~\ref{fig:huflex}. The solid line represents the observed
distribution, which is clearly skewed toward positive values. If
neither excess emission nor observational errors were present, the
distribution would be a $\delta$ function centered on 0. Therefore, we
consider the broadening toward negative values as a measure of the
combined photometric plus dereddening errors. By mirroring the
histogram for $L_\mathrm{F336W,0}$ about 0 (dashed histogram in
Figure~\ref{fig:huflex}) one can identify the stars with excess
emission as those populating the bins where the observed distribution
exceeds the mirrored one (hatched region). An inspection to
Figure~\ref{fig:huflex} shows that for
$L_\mathrm{F336W,exc}\gtrsim0.035~L_\sun$ the contamination due to
excesses induced by random errors becomes negligible.

These are, of course, the same stars with too blue a
$(m_\mathrm{F336W,0}-m_\mathrm{F439W,0})$ color for the
$(m_\mathrm{F439W,0}-m_\mathrm{F814W,0})$ one highlighted in
Figure~\ref{fig:ubbi}. As shown in Figure~\ref{fig:ubuvex}, the Balmer
continuum excess $L_\mathrm{F336W,0}$ displays a well defined
correlation with the $(m_\mathrm{F336W,0}-m_\mathrm{F439W,0})$ color,
albeit with a quite large scatter $\mathrm{rms}=0.04~L_{\sun}$. As
expected, the correlation is in the sense that the more negative the
color, the higher the inferred F336W excess. 

The excess emission was computed assuming for all of the stars a
metallicity appropriate for the young LMC population \citep[$Z=0.3\
Z_\sun$, e.g.][]{hil95,geha98}, in the sense that the observed F336W
fluxes were compared to models of that metallicity. This
simplification does not have any impact on our results, as we expect
the contamination from metal-poor ($Z\lesssim0.1\ Z_\sun$, \ie
intrinsically brighter in the F336W), field subgiants to be of the
order of only a few percent \citep{cole00}. In addition, the stellar
atmosphere models of \citet{bes98} indicate that the difference in the
F336W luminosity between $Z=0.3\ Z_\sun$ and $Z=0.1\ Z_\sun$ is of the
order of $0.01~L_\sun$ in the relevant range of effective temperature
($5,500~K\lesssim T_{eff}\lesssim8,000~K$, see
Figure~\ref{fig:hr_uvex}) and for $R_\ast=1.5\ R_\sun$, the typical
radius of these stars. This value is much smaller than the detection
threshold we have set at $L_\mathrm{F336W,exc}\gtrsim0.035~L_\sun$.
The models of \citet{bes98} also indicate that the effect of surface
gravity on photospheric emission in the F336W band is even smaller
than the one due to metallicity.

In conclusion, the 765 stars with an excess lager than about
$0.035~L_\sun$ are bona fide objects with F336W excess when compared to a
stellar photosphere.

\subsection{A sanity check: the correlation between U and {\boldmath $\ha$}
emission\label{sec:sanity}}
In addition to the U-band excess, Galactic Classical T~Tauri stars are also
observed to have \ha\ emission, which is also thought to be linked to
the accretion process from the circumstellar disk
\citep[e.g.][]{cal02}. As such, a correlation is to be expected
between these two quantities.

\begin{figure}
  \plotone{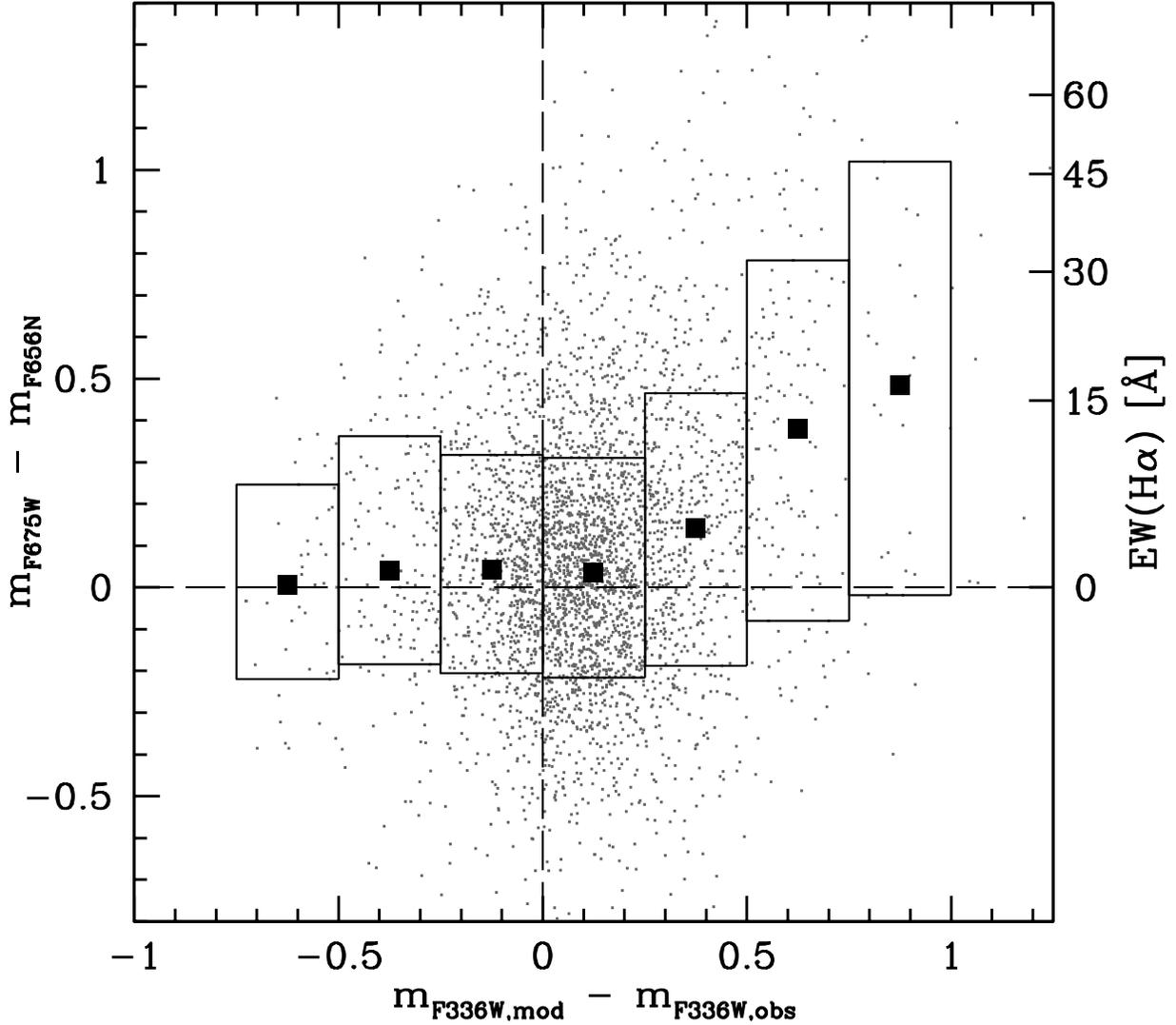}
  \caption{F336W vs \ha\ excess for our sample of candidate pre-Main
  Sequence stars in the field of SN~1987A (gray dots).
  $m_\mathrm{F336W,obs}$ is the observed magnitude,
  $m_\mathrm{F336W,mod}$ is the photospheric one from the models of
  \citet{bes98} and the $(m_\mathrm{F675W}-m_\mathrm{F656N})$
  color measures the \ha\ equivalent width \citep[e.g.][]{rom03}, as
  shown on the right vertical axis. The filled squares represent the
  median $(m_\mathrm{F675W}-m_\mathrm{F656N})$ value in the
  $(m_\mathrm{F336W,mod}-m_\mathrm{F336W,obs})$ bins marked by the
  rectangles, whose vertical extent includes 66\% of the stars in each
  bin. The high statistical correlation between the two quantities is
  apparent and is confirmed by the high value of Spearman's
  coefficient $\rho$ (see text). Note that this threshold also
  excludes stars with normal chromospheric activity \citep[3~\AA\ or
  $(m_\mathrm{F675W}-m_\mathrm{F656N})>0.15$,][]{fc94}.\label{fig:uv_ha}}
\end{figure}

In Figure~\ref{fig:uv_ha} we plot the excess emission in the F336W
filter versus the one in \ha: $m_\mathrm{F336W,obs}$ is the observed
magnitude, $m_\mathrm{F336W,mod}$ is the photospheric one from the
models of \citet{bes98} and the
$(m_\mathrm{F675W}-m_\mathrm{F656N})$ color measures the \ha\
equivalent width \citep[e.g.][]{rom03}. The correlation between U and
\ha\ activity is apparent.

A high statistical correlation between the two quantities is confirmed
by Spearman's coefficient $\rho$ \citep[see, e.g.,][]{con80}.  This is
a non-parametric test sensitive to correlations between two random
variables. The correlation coefficient is 7.9, which implies a
probability of less than $10^{-4}$ that the two variables are
uncorrelated.  Whereas the quality of the data does not allow to
derive the detailed \emph{shape} of the correlation, the existence of
the correlation itself is proven with a very high statistical
significance. This provides a very important sanity check as to the
pre-Main Sequence nature of these stars.

A potential source of concern here is chromospheric activity that
would produce excess emission both in the Balmer continuum \emph{and}
in \ha, without being linked to the presence of a circumstellar
accretion disk. As we will show in section~\ref{sec:pms_age}, the
stars we deal with in this paper have about $1~M_\sun$ (see
Figure~\ref{fig:age-mass}). The total chromospheric emission of a
solar-type star on the Subgiant Branch corresponds to a luminosity of
the order of $9.4\times 10^{-5}~L_\sun$ \citep{ulm79,pas00}. Even if
the flux were all concentrated in the spectral region covered by the
F336W filter, its contribution would still be negligible compared to
the threshold we have set of $3.5\times 10^{-2}~L_\sun$.  Moreover, it
is clear from Figure~\ref{fig:uv_ha} that all the objects with
significant F336W excess also have
$(m_\mathrm{F675W}-m_\mathrm{F656N})\gtrsim0.15$, \ie
$\mathrm{EW(\ha)}>3$, which is an upper limit to what can be
expected due to chromospheric activity \citep{fc94}.

In conclusion, then, the case under study the chromospheric emission
does not contribute in any significant way to the observed Balmer
continuum excess.

\subsection{From U-band excess to mass accretion\label{sec:uex2acc}}
Now that we have detected and characterized the stars with a statistically
significant excess in the U band we can compute the corresponding accretion 
rate. The following equation:

\begin{equation}
  L_{acc}\simeq\frac{GM_{\ast}\dot{M}}{R_{\ast}}\left(1-\frac{R_{\ast}}{R_{in}}
    \right)
  \label{eq:lacc_mdot}
\end{equation}
relates the bolometric accretion luminosity to the stellar and disk
parameters. $R_{in}$ is the inner radius of the accretion disk. Its
value is rather uncertain and depends upon the details of the coupling
of the accretion flow to the magnetic field. Following \citet{gull98}
we adopt $R_{in}=5\ R_{\ast}$ for all of the stars.

\citet{gull98} also provide an empirical relation between the accretion
luminosity $L_{acc}$ and the excess luminosity in the Johnson U passband
compared to a stellar photosphere:

\begin{equation}
   \log\left(\frac{L_{acc}}{L_\odot}\right)=1.09\ \log\left(
     \frac{L_{U,exc}}{L_\odot}\right)+0.98
  \label{eq:lacc_u}
\end{equation}
This, in turn, has been transformed to the WFPC2 F336W filter by
\citet{rob04} using theoretical models for the disk emission that
reproduce \citet{gull98}'s relation:

\begin{equation}
   \log\left(\frac{L_{acc}}{L_\odot}\right)=1.16\ \log
     \left(\frac{L_{\mathrm{F336W},exc}}{L_\odot}\right)+1.24
  \label{eq:lacc_ef336w}
\end{equation}

Combining equations~(\ref{eq:lacc_mdot}) and (\ref{eq:lacc_ef336w})
leads to a relation that links $\dot{M}$, the accretion rate in
$M_\odot\ \mathrm{yr}^{-1}$, to the excess luminosity in the F336W filter
$L_{\mathrm{F336W},exc}$:

\begin{equation}
  \log(\dot{M}) [M_\sun\ \mathrm{yr}^{-1}] =  -6.70
    +1.16\ \log\left(\frac{L_{\mathrm{F336W},exc}}{L_\sun}\right)
    + \log\left(\frac{R_\ast}
    {R_\odot}\right)-\log\left(\frac{M_\ast}{M_\odot}\right)
  \label{eq:mdot_ef336w}
\end{equation}

\subsection{The mass accretion rate\label{sec:acc}}
According to equation~(\ref{eq:mdot_ef336w}) in order to derive the
accretion rate {\boldmath $\dot{M}$} we need the following three
quantities:

\begin{itemize}
  \item {\boldmath $L_{\mathrm{F336W},exc}$} which is computed as described
in section~\ref{sec:uex} and whose resulting distribution is shown in
Figure~\ref{fig:huflex};

  \item the stellar radius {\boldmath $R_\ast$}, which is calculated
from the luminosity and temperature determined for every star using
the method developed by \citet{rom02}, \ie by fitting the observed
colors to the ones predicted by stellar model atmospheres, and
adopting a distance modulus to the LMC of 18.57 \citep{rom00};

  \item the stellar mass {\boldmath $M_\ast$} that we estimate by
comparing a star's location in the HR diagram with evolutionary tracks
computed for $Z=0.3\ Z_\sun$ \citep{sie97}.
\end{itemize}

The distribution of the derived accretion rates is shown in
Figure~\ref{fig:hmdot} for the stars with
$L_{\mathrm{F336W},exc}>0.035~L_\sun$, \ie those for which the excess
can be measured with high reliability. As it can be seen, the pre-Main
Sequence stars detected in the field of SN~1987A through their U-band
excess have accretion rates larger than $\simeq 1.5\times
10^{-8}M_\sun\ yr^{-1}$.

The apparent low cutoff at low values of $\dot{M}$ is not real, but,
rather, is a consequence of the threshold we have used to identify
stars with a significant Balmer continuum excess. In other words,
there might be stars in the field with smaller accretion rates, but
they would be lost in the measurement errors. Moreover, Galactic
T~Tauri stars are known to show variability at many wavelengths with
timescales as short as a few days \citep[see, for example, the review
by][]{ber89}. It is clear, then, that, since we only detect those
stars with the strongest excesses, our method tends to pick up stars
when they are the most active.

\section{Ages and masses of the pre-Main Sequence stars\label{sec:pms_age}}
In section~\ref{sec:uex} we noticed that a visual inspection to the HR
diagram indicates that, if interpreted as pre-Main Sequence objects,
the stars with U excess are consistent with an age of roughly
12~Myrs. We can now refine that estimate by comparing each star's
luminosity and temperature to those predicted as a function of mass
and age by the pre-Main Sequence evolutionary models of \citet{sie97}
for $Z=0.3\ Z_\sun$.

The resulting age distribution of the stars with
$L_{\mathrm{F336W},exc}>0.035~L_\sun$ is shown in the left panel of
Figure~\ref{fig:age-mass}. The peak of the distribution ($14\pm2$~Myr)
is in very good agreement with the age of the massive stars
($M>6~M_\sun$) in this same field \citep[$12\pm2$~Myrs,][]{pan00}.
The spread in the histogram is consistent with being produced by
photometric and dereddening errors only, which, for these stars, are
of the order of 20\% in the effective temperature (see also
Figure~\ref{fig:hr_uvex}). At this stage, then, we cannot assess if
the stars with significant U band excess show an intrinsic spread in
age or not. It is worth noticing that, if we take the comparison with
the \citet{sie97} isochrones at face value, almost no stars have ages
lower that about 5~Myrs.

The mass distribution of the same stars with significant Balmer
continuum excess is shown in the right panel of
Figure~\ref{fig:age-mass}. The mean mass of the accreting stars is
$1.25~M_\sun$, with an rms dispersion of $0.2~M_\sun$.

The mean value of the accretion rate for the stars with significant
Balmer continuum excess is $\simeq 2.5\times 10^{-8}M_\sun\ yr^{-1}$
(see Figure~\ref{fig:hmdot}). If we assume that this is a steady-state
mean value, then these stars still have to accrete a significant
fraction of their mass (10-20\%) before settling to their final Main
Sequence value. Of course, this is surely an overestimate for two main
reasons. As we have already noticed, we tend to pick up the stars when
their activity is maximum. In addition, accretion activity is observed
to slow down with time in Galactic pre-Main Sequence stars
\citep[e.g.][]{cal02}, breaking the simplistic steady-state
hypothesis.

\section{Summary and conclusions\label{sec:sum}}
We have identified about 800 stars with statistically significant
Balmer continuum excess in the field of SN~1987A in the Large
Magellanic Cloud. This excess positively correlates with \ha\
emission, as derived from the comparison of broad and narrow-band
photometry. Whereas the quality of the data does not allow to derive
the detailed \emph{shape} of the correlation, the existence of the
correlation itself is proven with a very high statistical significance
(Spearman's test returns a probability of less than $10^{-4}$ that the
two variables are uncorrelated). Both the Balmer continuum and the
\ha\ excesses are well above the levels expected from chromospheric
activity. These facts lead us to interpret these objects as pre-Main
Sequence stars.

In this framework, given their location in the HR diagram and the
emission in the Balmer continuum and \ha\ line, the objects in our
sample are the equivalent of Galactic Classical T~Tauri stars. This is
the first time that such a fundamental parameter as the accretion rate
is measured for this class of objects in a galaxy other than our own
Milky Way. Doing so in the Large Magellanic Cloud provides a unique
opportunity to sample astrophysical conditions not represented in
local star-forming regions, but that were common in the early
Universe.

When interpreted as pre-Main Sequence stars, the comparison of the
objects' location in the HR diagram with theoretical evolutionary
tracks allows one to derive their masses ($\sim1-1.4~M_\sun$) and ages
($\sim12-16$~Myrs). At such an age and with an accretion rate in
excess of $\sim 1.5\times10^{-8}M_\sun\ yr^{-1}$, these candidate
pre-Main Sequence stars in the field of SN~1987A are both older and
more active than their Galactic counterparts known to date. In fact,
the overwhelming majority of T~Tauri stars in Galactic associations
seem to dissipate their accretion disks before reaching an age of
about 6~Myrs \citep{hai01,arm03}. Moreover, the oldest Classical
T~Tauri star know in the Galaxy, \objectname{TW~Hydr\ae} at an age of
10~Myrs, \ie comparable to that of our sample stars, has a measured
accretion rate some 30 times lower than the stars in the neighborhood
of SN~1987A \citep{muz00}.

\begin{figure}
  \plotone{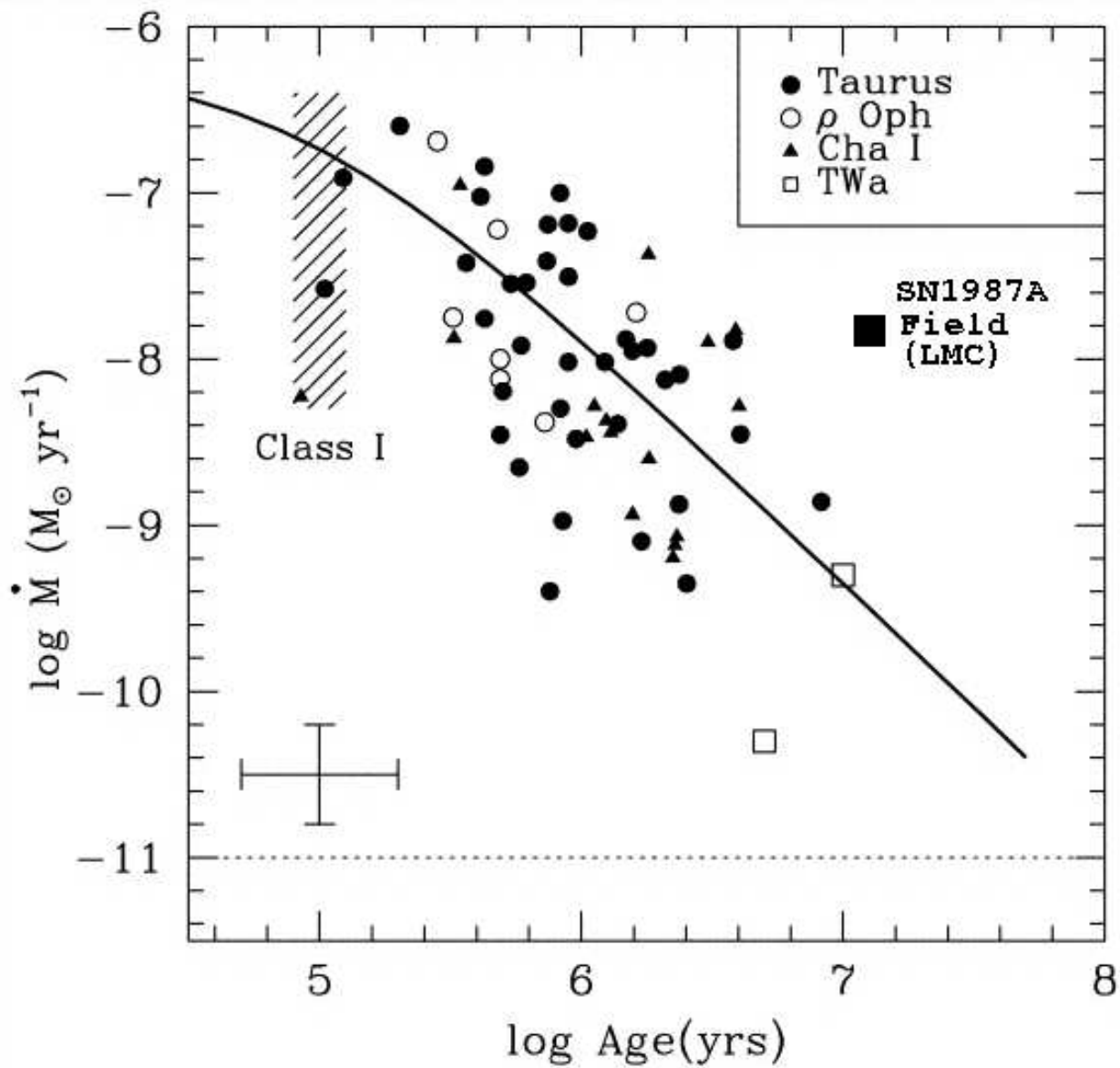}
   \caption{Mass accretion rate as a function of age for Classical
   T~Tauri stars in different star-forming regions \citep[adapted
   from][]{muz00}.  The solid line represents the fiducial model from
   the viscous disk similarity solutions of \citet{har98}. The
   errorbar indicates, for the Galactic stars, typical uncertainties
   in ages from the HR diagram and typical uncertainties in accretion
   rates due to variability. Our result for the field of SN~1987A in
   the LMC is marked with a black square. Its uncertainties are
   discussed in section~\ref{sec:sum}.
   \label{fig:muz}}
\end{figure}

The situation is summarized in Figure~\ref{fig:muz}, adapted from
\citet{muz00}, where we compare the position in the age-$\dot{M}$
plane of the stars described in this paper with that of members of
Galactic star-forming regions. An obvious selection bias that affects
our census is that we only detect those stars with the largest Balmer
continuum excesses, \ie highest accretion rates. There might be stars
in the field with smaller accretion rates, either intrinsically or
because they were observed when the accretion activity was at a
minimum, which fall below our detection threshold. This selection
effect is rather hard to quantify, but it is clear that the locus of
the accreting stars that we do detect in the neighborhood of SN~1987A
is significantly displaced from the one defined by local pre-Main
Sequence stars.

There are essentially two ways to reconcile the position of the LMC
point in Figure~\ref{fig:muz} with the mean locus observed in the
Galaxy: significantly reduce the age of the stars in the SN~1987A
region and/or significantly decrease their inferred accretion rate.
Neither option, however, seems viable. In fact, while the
uncertainties on either quantity can be quite large for any given
star, the LMC point reflects the mean values of several hundred stars,
thus making the random errors, for all practical purpose,
negligible. Let us now consider the systematic errors.

\begin{figure}
  \plottwo{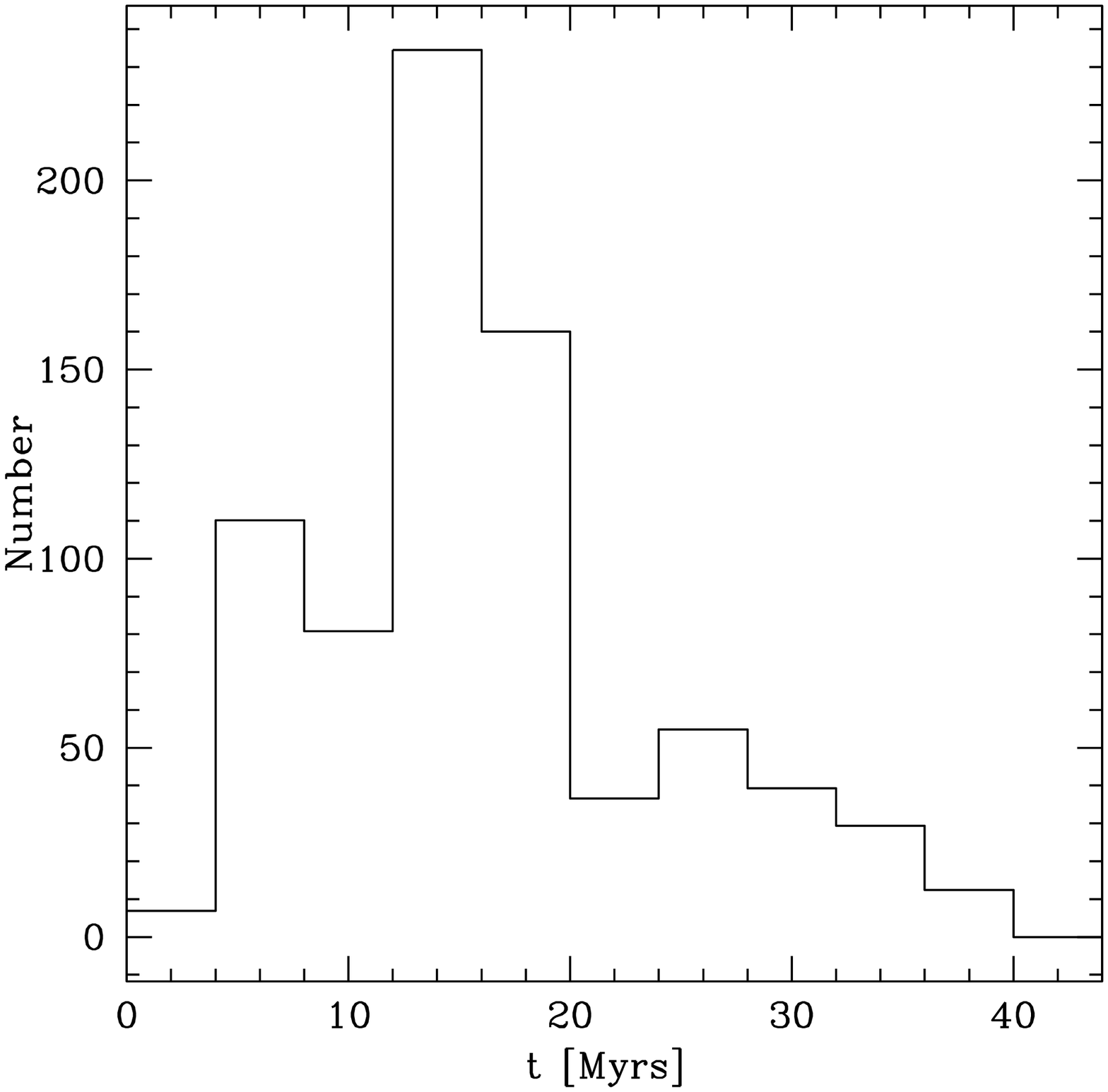}{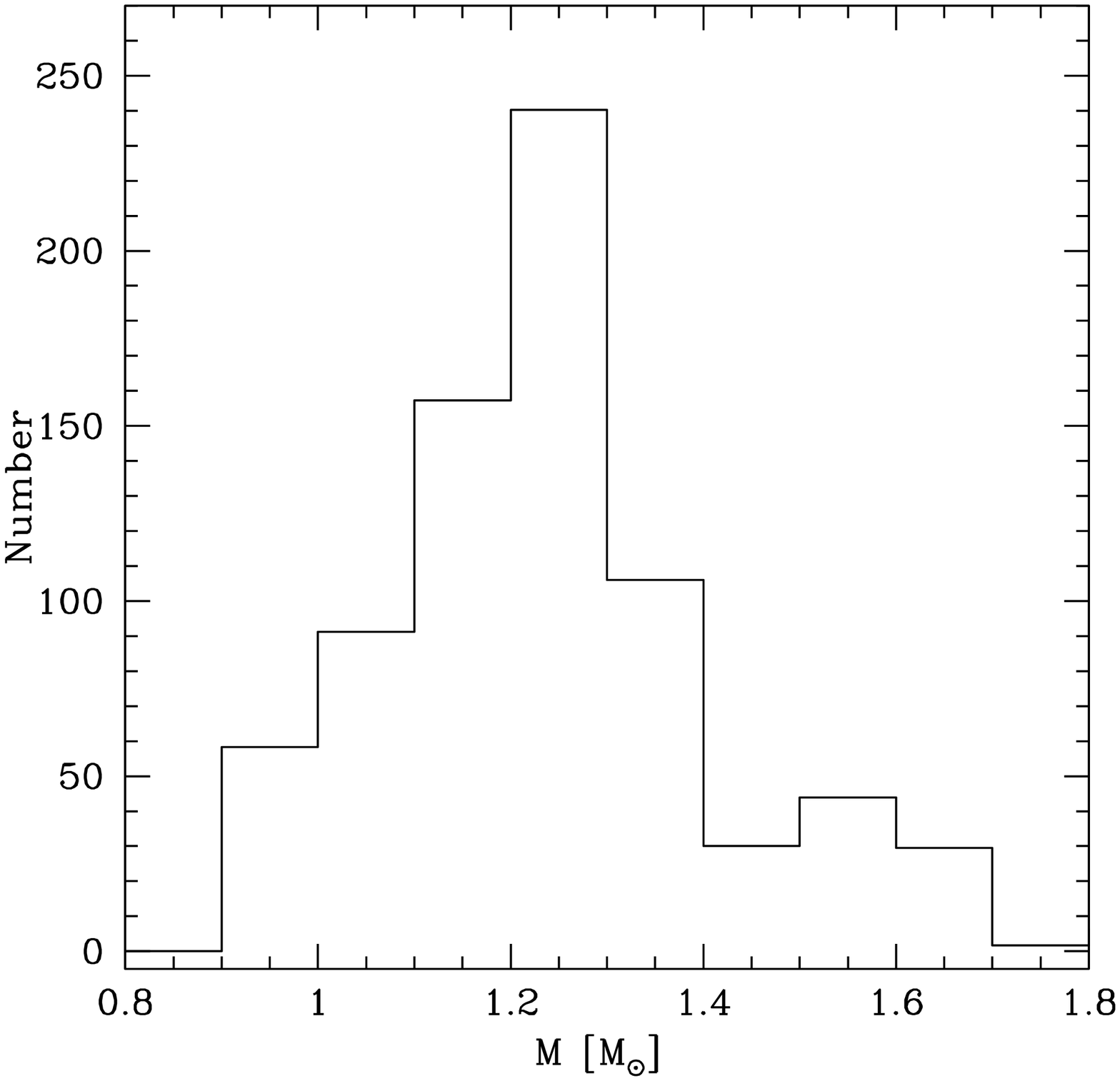}
  \caption{Age (\emph{left panel}) and mass (\emph{right panel})
  distributions of the stars with highly reliable U-band excess
  ($L_{\mathrm{F336W},exc}>0.035~L_\sun$).
  Both quantities were derived by comparing the stars' location in the
  HR diagram to the theoretical pre-Main Sequence evolutionary tracks
  for $Z=0.3\ Z_\sun$ of \citet{sie97}. The luminosities and
  temperatures of the stars were computed excluding the F336W
  magnitude from the fit, since it is affected by non-photospheric emission.
  \label{fig:age-mass}}
\end{figure}

If we move the LMC point \emph{horizontally} in Figure~\ref{fig:muz},
in order to occupy the same area as Galactic stars, the LMC stars
would have to be younger than about 4~Myrs or less, the age of the
oldest Galactic star with a measured accretion rate of
$1.5\times10^{-8}M_\sun\ yr^{-1}$. This would require our estimate
of the ages of the LMC stars to be systematically wrong by a factor of
three or more. Such a large shift is rather implausible on accounts of
two main considerations. First, \citet{rom02} have shown that the
dereddening technique we have adopted here does not introduce any
significant systematic errors on the stellar parameter, hence, given a
set of pre-Main Sequence evolutionary tracks, on the derived
age. Second, the age of 14~Myrs inferred from the peak of the
histogram in Figure~\ref{fig:age-mass} agrees very well both with that
of the most massive stars in the field \citep[$12\pm2$~Myrs,][]{pan00}
and that of the progenitor of SN~1987A \citep[10-12~Myrs see, for
example,][and references therein]{scud96}.  These age estimates are,
of course, completely independent from that of the low-mass stars we
are discussing here, lending support to it.

\begin{figure}
  \plotone{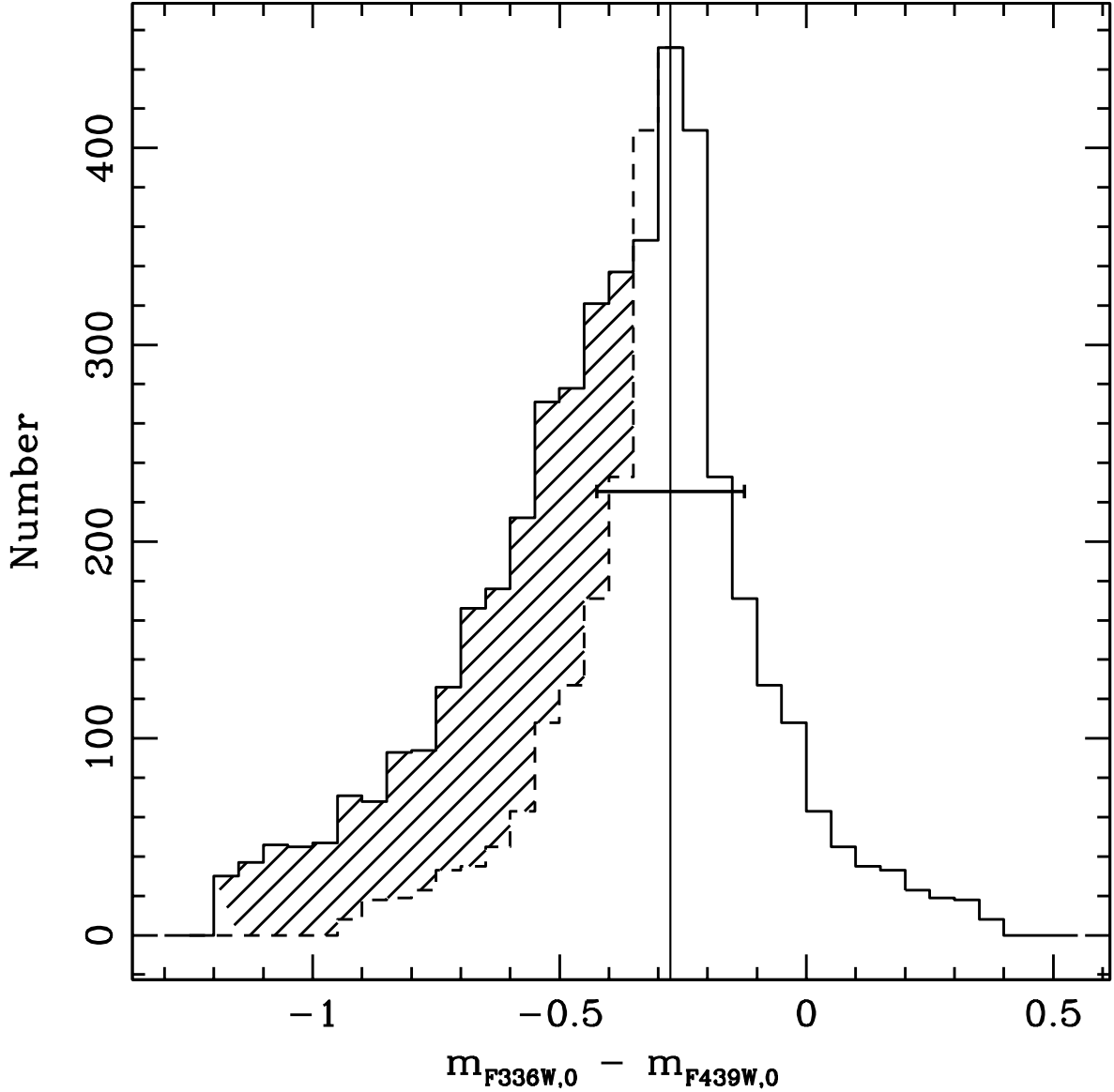}
  \caption{Distribution of $(m_\mathrm{F336W,0}-m_\mathrm{F439W,0})$
  color for stars with
  $0.6<(m_\mathrm{F439W,0}-m_\mathrm{F814W,0})<1.2$ (see
  Figure~\ref{fig:ubbi}). The full line is the observed histogram
  while the dashed one is the reflection of its red part about the
  peak (thin vertical line), which also corresponds to the expected
  color of the turnover point in Figure~\ref{fig:ubbi}. The asymmetry
  of the $(m_\mathrm{F336W,0}-m_\mathrm{F439W,0})$ toward blue colors,
  \ie F336W excess, is apparent and is also highlighted by the hatched
  area. The FWHM of the distribution expected from the measured
  photometric and dereddening errors on the
  $(m_\mathrm{F336W,0}-m_\mathrm{F439W,0})$ color is shown as an
  horizontal errorbar. As it can be seen, it nicely reproduces the
  part of the histogram to the right of the peak, while it severely
  underestimates the width of one to the left.\label{fig:hubex}}
\end{figure}

\begin{figure}
  \plotone{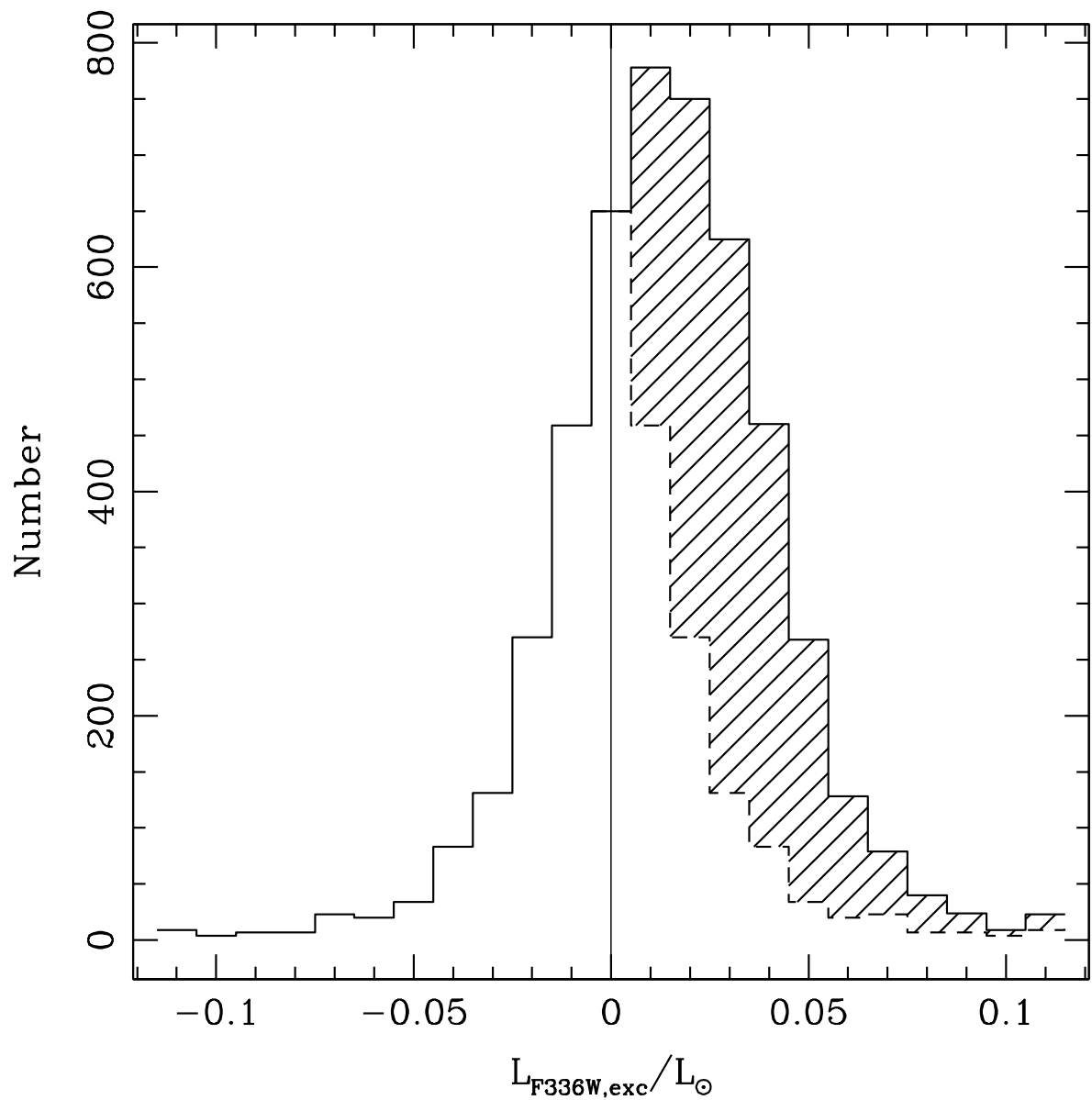}
  \caption{Histogram of the measured flux excess in the F336W filter
    (full line). The stars with an excess emission statistically beyond
    the observational error are highlighted in the hatched area (see text).
    The contamination due to spurious emission becomes negligible above
    about $0.035~L_\sun$ and we set this as our detection threshold.
    \label{fig:huflex}}
\end{figure}

\begin{figure}
  \plotone{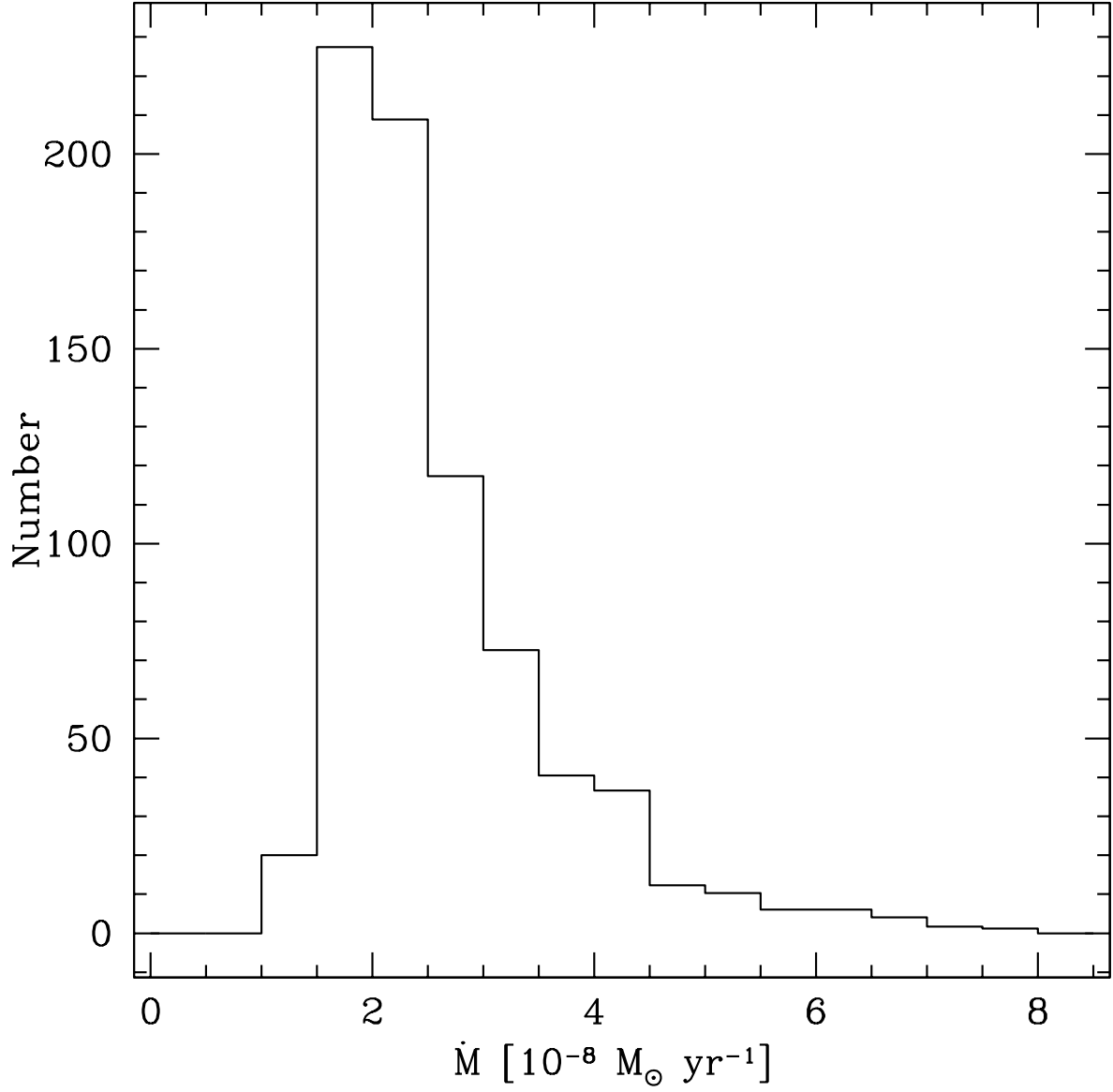}
  \caption{Accretion rates as derived from the excess flux in the
    F336W filter (see equation~\ref{eq:mdot_ef336w}) for the stars for
    which the excess can be measured with certainty
    ($L_\mathrm{F336W,exc}>0.035~L_\sun$, see
    Figure~\ref{fig:huflex}).\label{fig:hmdot}}
\end{figure}

The other possibility is that, given the age of about 12-16~Myrs, we
have overestimated the accretion rate onto the candidate pre-Main
Sequence stars in the surroundings of SN~1987A. A factor of about 30
reduction is required in order to make our value for the LMC to agree
with TW~Hydr\ae, the oldest accreting T~Tauri known in our Galaxy (see
Figure~\ref{fig:age-mass}). According to
equation~(\ref{eq:mdot_ef336w}) this corresponds to reducing the
Balmer continuum excess $L_{\mathrm{F336W},exc}$ by a factor of almost
20 (for comparison, this would be so small as to be even smaller than
the bin size of the histogram in Figure~\ref{fig:huflex}). Such a
reduction, however, is clearly incompatible with the observed
distribution of the excesses, in particular with its asymmetry, which
marks the stars with a Balmer continuum excess beyond statistical
errors (see also Figure~\ref{fig:hubex}).

\begin{figure}
  \plotone{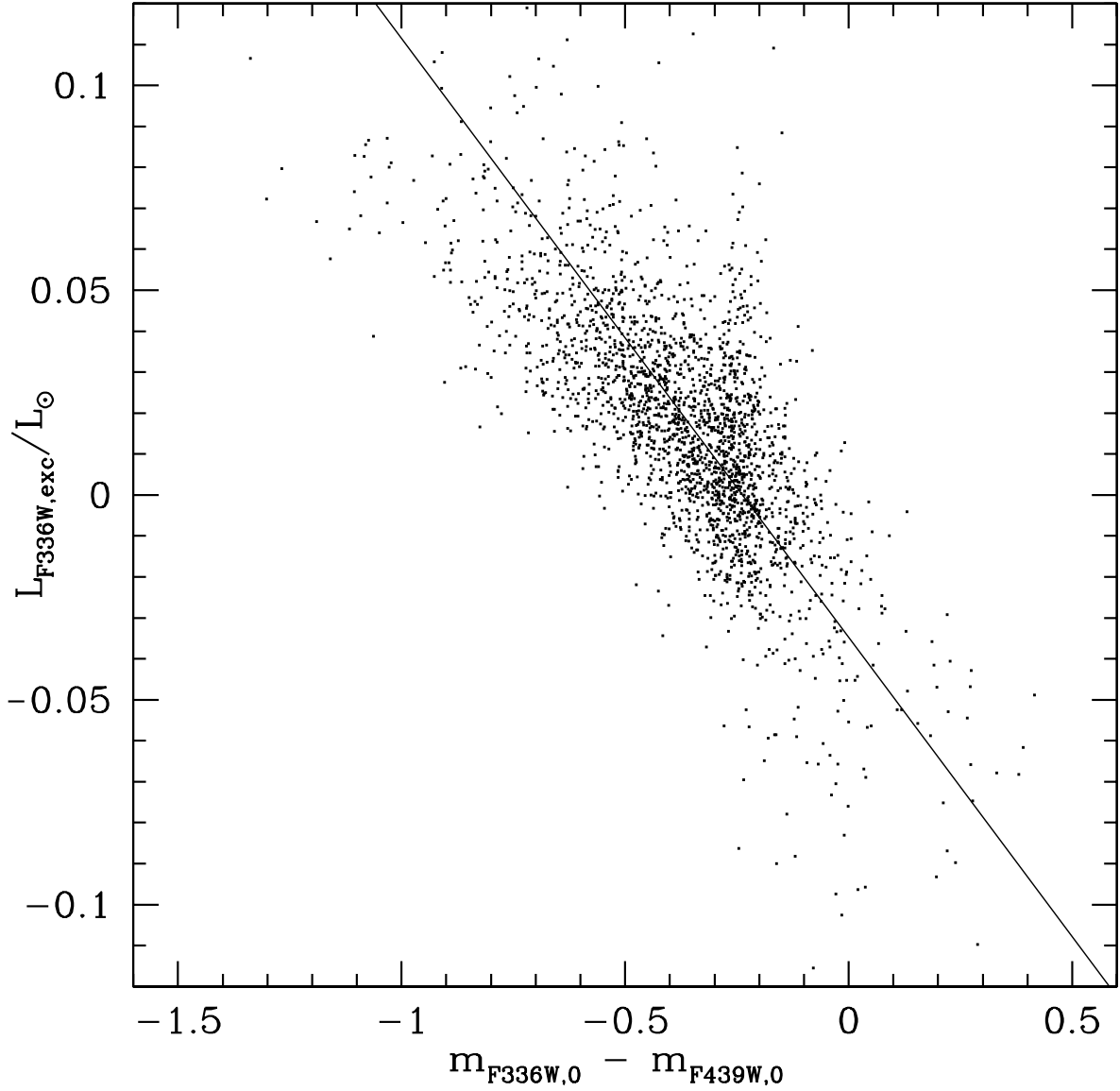}
  \caption{Balmer continuum excess vs dereddened
    $(m_\mathrm{F336W,0}-m_\mathrm{F439W,0})$ color for the stars with
    good overall photometry ($\bar{\delta}_5<0.1$). The result of a
    linear regression is displayed as a solid line. As expected, the
    two quantities show a well defined correlation.\label{fig:ubuvex}}
\end{figure}

A potential source of concern is the effect on the derived accretion
rate of the presence of circumstellar dust in the immediate
surroundings of the candidate pre-Main Sequence stars (flared disks,
remnants of the cocoon they were formed out of, etc.). This extra
source of reddening is not accounted for by our method and the
candidate pre-Main Sequence stars could be affected by a higher value
of $E(B-V)$ than mean one from the 4 closest neighbors with a direct
determination, \ie have bluer intrinsic
$(m_\mathrm{F336W,0}-m_\mathrm{F439W,0})$ colors. According to the
data in Figure~\ref{fig:ubuvex}, however, this would correspond to
increasing the inferred Balmer continuum excess and, consequently, the
accretion rate, further confirming the difference with what is
observed in the Galaxy.

Finally, let us notice that the accretion rate for several of the
Galactic stars in Figure~\ref{fig:age-mass} was derived with the same
technique we have used in this paper, \ie the relation between Balmer
continuum excess and accretion rate by \citet{gull98}. As such, they
are affected by the same issues on the calibration of this relation as
the LMC stars presented here (if any). The discrepancy between
Galactic stars and those in the neighborhood of SN~1987A in the LMC,
then, appears to be real.  Of course, spectroscopic data on these
stars would be highly desirable to better characterize their
properties.

A higher accretion rate in the LMC than in the Galaxy was also
suggested for A and B spectral type pre-Main Sequence stars by
\citet{lam99} and \citet{dewit02}. The combination of this result with
ours points towards a significantly higher accretion activity over a
factor of about 6 in mass for pre-Main Sequence stars at lower
metallicity.

\acknowledgments
We warmly thank Luca Pasquini for enlightening discussions on stellar
chromospheric activity. The detailed comments and suggestions from an
anonymous referee are gratefully acknowledged.

\clearpage

\clearpage

\clearpage

\clearpage

\clearpage

\clearpage

\clearpage

\clearpage

\clearpage

\clearpage

\clearpage

\begin{table*}[!ht]
\begin{center}
\caption{Log of the observations centered on 
Supernova~1987A.}
\begin{tabular}{*{5}{c}}
& & & & \\
Filter Name &  \multicolumn{3}{c}{Exposure Time (seconds)}  & Comments\\
\cline{2-4} &September 1994\tablenotemark{a}
&February 1996\tablenotemark{b}
&July 1997\tablenotemark{c}   & \\ \tableline
{\bf F255W} &      2x900      &    1100+1400   &   2x1300    & UV~Filter  \\
{\bf F336W} &      2x600      &      2x600     &   2x800     & ~U~Filter  \\
{\bf F439W} &      2x400      &     350+600    &   2x400     & ~B~Filter  \\
{\bf F555W} &      2x300      &      2x300     &   2x300     & ~V~Filter  \\
{\bf F675W} &      2x300      &      2x300     &   2x300     & ~R~Filter  \\
{\bf F814W} &      2x300      &      2x300     &   2x400     & ~I~Filter
\\[0.2cm]\tableline & & & & \\[-0.4cm]
{\bf F656N} &      ---        & 1100+1300\tablenotemark{d} &  4x1400 & \ha \\
\end{tabular}
\end{center}

\tablenotetext{a}{September 24, 1994, proposal number 5753 (SINS collaboration,
PI R.P. Kirshner).}
\tablenotetext{b}{February 6, 1996, proposal number 6020 (SINS collaboration,
PI R.P. Kirshner).}
\tablenotetext{c}{July 10, 1997, except for F502N taken on
July 12, 1997, proposal number 6437 (SINS collaboration, PI R.P. Kirshner).}
\tablenotetext{d}{February 3, 1994 proposal number 5203 (PI J. Trauger).}
\label{tab:log}
\end{table*}

\end{document}